\documentclass[12pt]{article}
\usepackage{a4wide,epsfig,amsmath}

\newcommand{\agt}{\rlap{\lower 3.5 pt \hbox{$\mathchar \sim$}} \raise 1pt
 \hbox {$>$}}
\newcommand{\alt}{\rlap{\lower 3.5 pt \hbox{$\mathchar \sim$}} \raise 1pt
 \hbox {$<$}}
\newcommand{\Li}{\mathop{\mathrm{Li}}\nolimits}
\newcommand{\arcosh}{\mathop{\mathrm{arcosh}}\nolimits}
\newcommand{\artanh}{\mathop{\mathrm{artanh}}\nolimits}

\catcode`@=11
\newcount\@tempcntc
\def\@citex[#1]#2{\if@filesw\immediate\write\@auxout{\string\citation{#2}}\fi
  \@tempcnta\z@\@tempcntb\m@ne\def\@citea{}\@cite{\@for\@citeb:=#2\do
    {\@ifundefined
       {b@\@citeb}{\@citeo\@tempcntb\m@ne\@citea\def\@citea{,}{\bf
?}\@warning
       {Citation `\@citeb' on page \thepage \space undefined}}%
    {\setbox\z@\hbox{\global\@tempcntc0\csname b@\@citeb\endcsname\relax}%
     \ifnum\@tempcntc=\z@ \@citeo\@tempcntb\m@ne
       \@citea\def\@citea{,}\hbox{\csname b@\@citeb\endcsname}%
     \else
      \advance\@tempcntb\@ne
      \ifnum\@tempcntb=\@tempcntc
      \else\advance\@tempcntb\m@ne\@citeo
      \@tempcnta\@tempcntc\@tempcntb\@tempcntc\fi\fi}}\@citeo}{#1}}
\def\@citeo{\ifnum\@tempcnta>\@tempcntb\else\@citea\def\@citea{,}%
  \ifnum\@tempcnta=\@tempcntb\the\@tempcnta\else
   {\advance\@tempcnta\@ne\ifnum\@tempcnta=\@tempcntb \else
\def\@citea{--}\fi
    \advance\@tempcnta\m@ne\the\@tempcnta\@citea\the\@tempcntb}\fi\fi}
\catcode`@=12

\begin{document}

\title{
\vskip-3cm{\baselineskip14pt
\centerline{\normalsize DESY 12--033\hfill ISSN 0418-9833}
\centerline{\normalsize March 2012\hfill}}
\vskip1.5cm
Bottom-flavored hadrons from top-quark decay at next-to-leading order in the
general-mass variable-flavor-number scheme}

\author{Bernd A. Kniehl, Gustav Kramer\\
{\normalsize II. Institut f\"ur Theoretische Physik, Universit\"at Hamburg,}\\
{\normalsize Luruper Chaussee 149, 22761 Hamburg, Germany}\\
\\
Seyed M. Moosavi Nejad\\
{\normalsize Faculty of Physics, Yazd University, P.O. Box 89195--741, Yazd,
IRAN,}\\
{\normalsize School of Particles, Institute for Research in Fundamental
Sciences (IPM)},\\
{\normalsize P.O. Box 19395--5746, Tehran, Iran}
}

\date{}

\maketitle

\begin{abstract}
We study the scaled-energy ($x_B$) distribution of bottom-flavored hadrons
($B$) inclusively produced in top-quark decays at next-to-leading order (NLO)
in the general-mass variable-flavor-number scheme endowed with realistic,
nonperturbative fragmentation functions that are obtained through a global fit
to $e^+e^-$ data from CERN LEP1 and SLAC SLC exploiting their universality and
scaling violations.
Specifically, we study the effects of gluon fragmentation and finite
bottom-quark and $B$-hadron masses.
We find the NLO corrections to be significant.
Gluon fragmentation leads to an appreciable reduction in the partial decay
width at low values of $x_B$.
Hadron masses are responsible for the low-$x_B$ threshold, while the
bottom-quark mass is of minor importance.
Neglecting the latter, we also study the doubly differential distribution
$d^2\Gamma/(dx_B\,d\cos\theta)$ of the partial width of the decay
$t\to bW^+\to B\ell^+\nu_\ell+X$, where $\theta$ is the decay angle of the
charged lepton in the $W$-boson rest frame.
\medskip

\end{abstract}

\newpage

\section{Introduction}

Among other things, the CERN Large Hadron Collider (LHC) is a superlative top
factory, producing about 90 million top-quark pairs per year of running at
design energy $\sqrt{S}=14$~TeV and design luminosity
${\cal L}=10^{34}$~cm$^{-2}$s$^{-1}$ in each of the four experiments
\cite{Moch:2008qy}.
This will allow us to determine the properties of the top quark, such as its
mass $m_t$, total decay width $\Gamma_t$, branching fractions, and elements
$V_{tq}$ of the Cabibbo-Kobayashi-Maskawa (CKM) \cite{Cabibbo:1963yz} quark
mixing matrix, with unprecedented precision.
Due to its large mass, the top quark decays so rapidly that it has no time to
hadronize and passes on its full spin information to its decay products.  
If it were not for the confinement of color, the top quark could, therefore,
be considered as a free particle.
Due to $|V_{tb}|\approx1$, top quarks almost exclusively decay to bottom
quarks, via $t\to bW^+$.

On the other hand, bottom quarks hadronize, via $b\to B+X$, before they decay,
so that the decay process $t\to BW^++X$ is of prime importance, and it is an
urgent task to predict its partial decay width as realistically and reliably
as possible.
Of particular interest are the distribution in the scaled $B$-hadron energy
$x_B$ in the top-quark rest frame, and, in the case of leptonic $W$-boson
decays $W^+\to\ell^+\nu_\ell$, the one in the charged-lepton decay angle 
$\theta$ in the $W$-boson rest frame.
In fact, the $x_B$ distribution provides direct access to the  $B$-hadron
fragmentation functions (FFs), and the $\cos\theta$ distribution allows us to
analyze the $W$-boson polarization and so to further constrain the $B$-hadron
FFs by exploiting $x_B$ distributions for all the $W$-boson polarization
states. 
$B$ mesons are, for instance, cleanly identified by their decays to $J/\psi$
mesons, which are easy to tag through their spectacular decays to $e^+e^-$ and
$\mu^+\mu^-$ pairs.

The theoretical aspects of top-quark physics at the LHC are nicely summarized
in a recent review paper \cite{Bernreuther:2008ju}.
In the approximation of treating the bottom quark as a stable final-state
particle that does not hadronize, the QCD corrections to $\Gamma(t\to bW^+)$
are known at NLO with subsequent $W^+\to\ell^+\nu_\ell$ decay
\cite{Jezabek:1988iv,Jezabek:1988ja} and at next-to-next-to-leading order for
stable $W$ boson \cite{Czarnecki:1998qc}.
The terms of order $\beta_0^n \alpha_s^{n+1}$, where $\beta_0$ is the first
coefficient of the QCD beta function and $\alpha_s$ is the strong-coupling
constant, were resummed to all orders in Ref.~\cite{Mehen:1997mw}.
The NLO electroweak corrections were found in Ref.~\cite{Denner:1990ns}.

The hadronization of the bottom quark was considered in the NLO QCD analyses
of top-quark decay in
Refs.~\cite{Corcella:2001hz,Cacciari:2002re,Corcella:2009rs} and was,
in fact, identified to be the largest source of uncertainty in the
determination of the top-quark mass.
In Refs.~\cite{Corcella:2001hz,Cacciari:2002re,Corcella:2009rs}, the $W$ boson
was taken to be stable, the bottom-quark mass was neglected at the parton
level, and the hadronization process $b\to B+X$ was implemented as a
convolution of a perturbative FF \cite{Mele:1990cw}, describing in a way the
conversion of the massless bottom quark to its massive counterpart, and a
nonperturbative FF modeling the hadronization of the latter.
Hereby, the perturbative FF depends on a factorization scale and is subject to
Dokshitzer-Gribov-Lipatov-Altarelli-Parisi (DGLAP) \cite{Gribov:1972ri}
evolution, while the nonperturbative FF is scale independent.
Because of the treatment of a heavy quark as a massless parton, this framework
corresponds to the zero-mass variable-flavor-number (ZM-VFN) scheme.
In Refs.~\cite{Corcella:2001hz,Cacciari:2002re}, also soft-gluon resummation
was studied.
In Ref.~\cite{Corcella:2009rs}, also the distribution in the invariant mass
$m_{B\ell}$ of the $B$ hadron and the charged lepton $\ell$ from $W$-boson
decay was considered.

In this paper, we revisit $B$-hadron production from top-quark decay working
at NLO in the general-mass variable-flavor-number (GM-VFN) scheme, which was
elaborated for inclusive heavy-flavored-hadron production in $e^+e^-$
annihilation \cite{Kneesch:2007ey}, two-photon collisions \cite{Kramer:2001gd},
photoproduction \cite{Kramer:2003jw}, and hadroproduction
\cite{Kniehl:2004fy,Kniehl:2008zza}, and provides an ideal theoretical
framework also here.
Being manifestly based on Collin's hard-scattering factorization theorem
appropriate for massive quarks \cite{Collins:1998rz}, this factorization scheme
allows one to resum the large logarithms in $m_b$, to retain the finite-$m_b$
effects, and to preserve the universality of the FFs, whose scaling violations
remain to be subject to DGLAP evolution.
In this way, it combines the virtues of the ZM-VFN and fixed-flavor-number
(FFN) schemes and, at the same time, avoids their flaws.
It is, in fact, a tailor-made tool for global analyses of experimental data on
the inclusive production of heavy-flavored hadrons, allowing one to transfer
nonperturbative information on the hadronization of quarks and gluons from one
type of experiment to another and, within one type of experiment, from one
energy scale $\mu_F$ to another, without the restriction $\mu_F\gg m_b$
inherent to the ZM-VFN scheme.
In the GM-VFN scheme, the perturbative FFs enter the formalism via subtraction
terms for the hard-scattering cross sections and decay rates, so that the
actual FFs are truly nonperturbative and may be assumed to have some smooth
forms that can be determined through global data fits.  
In contrast to the FFN scheme, the GM-VFN scheme also accommodates FFs for
gluons and light quarks, as in the ZM-VFN scheme.

Specifically, our analysis is supposed to enhance those of
Refs.~\cite{Corcella:2001hz,Cacciari:2002re,Corcella:2009rs} by retaining all
nonlogarithmic $m_b$ terms of the result in the FFN scheme and by including
light-parton fragmentation.
Furthermore, we include finite-$m_B$ effects, which modify the relations
between partonic and hadronic variables and reduce the available phase space.
Although these additional effects are not expected to be truly sizable
numerically, except for certain corners the phase space, their study is
nevertheless mandatory in order to fully exploit the enormous statistics of
the LHC data to be taken in the long run for a high-precision determination of
the top-quark properties.
We also extend Refs.~\cite{Corcella:2001hz,Cacciari:2002re,Corcella:2009rs} by
including subsequent leptonic decays of the $W$ boson and studying the
distribution in the angle $\theta$ of the charged lepton in the $W$-boson rest
frame, while Ref.~\cite{Corcella:2009rs} is focused on the $m_{B\ell}$
distribution.

This paper is organized as follows.
In Sec.~\ref{sec:one}, we explain how to incorporate finite-$m_B$ corrections
in the evaluation of $d\Gamma(t\to B+X)/d x_B$.
In Sec.~\ref{sec:two}, we give the parton-level expressions for
$d\Gamma(t\to B+X)/dx_B$ at NLO in the ZM-VFN and FFN schemes and combine
them to obtain those in the GM-VFN scheme.
In Sec.~\ref{sec:three}, we list the parton-level formulas needed to evaluate
$d^2\Gamma(t\to B\ell\nu_\ell+X)/(dx_B\,d\cos\theta)$ at NLO in the ZM-VFN
scheme.
In Sec.~\ref{sec:four}, we present our numerical analysis.
In Sec.~\ref{sec:five}, we summarize our conclusions.
The Appendix accommodates some formulas that are too lengthy to be displayed
in Sec.~\ref{sec:three}.

\section{Hadron mass effects}
\label{sec:one}

We consider the decay process
\begin{equation}
t(p_t)\to b(p_b)+W^+(p_W)(+g(p_g))\to B(p_B)+X,
\label{eq:proc}
\end{equation}
where $X$ collectively denotes the unobserved final-state particles and the
four-momentum assignments are indicated in parentheses.
The gluon in Eq.~(\ref{eq:proc}) contributes to the real radiation at NLO.
Both the $b$ quark and the gluon may hadronize to the $B$ hadron.

In the top-quark rest frame, the $b$ quark, gluon, and $B$ hadron have energies
$E_i=p_t\cdot p_i/m_t$ ($i=b,g,B$), which range from
$E_b^\text{min}=m_b$, $E_g^\text{min}=0$, and $E_B^\text{min}=m_B$ to
$E_b^\text{max}=(m_t^2+m_b^2-m_W^2)/(2m_t)$,
$E_g^\text{max}=[m_t^2-(m_b+m_W)^2]/(2m_t)$,
and $E_B^\text{max}=(m_t^2+m_B^2-m_W^2)/(2m_t)$, respectively.
In the case of gluon fragmentation, $g\to B$, the maximum $B$-hadron energy is
$\tilde{E}_B^\text{max}=[m_t^2+m_B^2-(m_b+m_W)^2]/(2m_t)$.
It is convenient to introduce the scaled energies
$x_i=E_i/E_b^\text{max}$ ($i=b,g,B$).

We wish to calculate the partial decay width of process~(\ref{eq:proc})
differential in $x_B$, $d\Gamma/dx_B$, at NLO in the GM-VFN scheme taking into
account finite-$m_B$ corrections.
In the ZM-VFN scheme, the four-momenta of the produced hadron and the mother
parton are related as $p_B=zp_a$ ($a=b,g$), where the scaling variable $z$
takes the values $0\le z\le1$.
This simple relation is not compatible with finite quark and/or hadron masses
and needs to be generalized when passing from the ZM-VFN scheme to the GM-VFN
scheme.
There is some freedom in defining the scaling variable in the presence of quark
and/or hadron masses.
In the case under consideration, a convenient choice is $E_B=zE_a$
\cite{Kneesch:2007ey}, {\it i.e.}\ to retain just one of the four equations
$p_B=zp_a$.
By the factorization theorem of the QCD-improved parton model, we then have
\begin{equation}
d\Gamma=\sum_{a=b,g}\int_0^1dz\,\left.\vphantom{\frac{1}{1}}
d\hat{\Gamma}_a(\mu_R,\mu_F)\right|_{E_a=E_B/z}D_a(z,\mu_F),
\label{eq:fact}
\end{equation}
where $d\hat{\Gamma}_a(\mu_R,\mu_F)$ is the differential decay width of the
parton-level process $t\to a+X$, with $X$ comprising the $W$ boson and any
other parton, and $D_a(z,\mu_F)$ is the FF of the transition $a\to B$.
Here, $\mu_R$ and $\mu_F$ are the renormalization and factorization scales,
respectively.
Substituting $d\hat{\Gamma}=dx_B(dx_a/dx_B)d\hat{\Gamma}/dx_a$ and eliminating
$z=x_B/x_a$ as the integration variable, we obtain our master formula
\begin{equation}
\frac{d\Gamma}{dx_B}=\sum_{a=b,g}\int_{x_a^\text{min}}^{x_a^\text{max}}
\frac{dx_a}{x_a}\,\frac{d\hat{\Gamma}_a}{dx_a}(\mu_R,\mu_F)
D_a\left(\frac{x_B}{x_a},\mu_F\right).
\label{eq:master}
\end{equation}
Using $x_B\le x_a$ along with the above bounds on $E_i$ ($i=b,g,B$), we have
$x_b^\text{min}=\max(\rho_b,x_B)$,
$x_b^\text{max}=1$,
$x_g^\text{min}=x_B$, and 
$x_g^\text{max}=\rho$,
where $\rho_i=m_i/E_b^\text{max}$ ($i=b,g,B$) and
$\rho=E_g^\text{max}/E_b^\text{max}$.
The kinematically allowed $x_B$ ranges are
$\rho_B\le x_B\le\min(1,E_B^\text{max}/E_b^\text{max})$ for $a=b$ and
$\rho_B\le x_B\le\rho$ for $a=g$.
In reality, we have $m_b<m_B$, so that $x_b^\text{min}=x_B$ and
$x_B^\text{max}=1$ for $a=b$.

In order to assess the theoretical uncertainty due to the freedom in the
choice of scaling variable in the GM-VFN scheme, we also consider here the
definition in terms of the plus component $V^+=(V^0+V^3)/\sqrt2$ of a
four-vector $V$ in light-cone coordinates.
Taking the three-axis to point along the common flight direction of parton $a$
and the $B$ hadron, we define $p_B^+=z p_a^+$ \cite{Albino:2005gd}.
This definition is invariant under boosts along the three-axis.
Starting from the factorization formula~(\ref{eq:fact}) with $E_a=E_B/z$
replaced by $p_a^+=p_B^+/z$, we obtain
\begin{equation}
\frac{d\Gamma}{dx_B}=\frac{1}{\sqrt{x_B^2-\rho_B^2}}
\sum_{a=b,g}\int_{x_a^\text{min}}^{x_a^\text{max}}
dx_a\,z\frac{d\hat{\Gamma}_a}{dx_a}(\mu_R,\mu_F)D_a\left(z,\mu_F\right),
\label{eq:lightcone}
\end{equation}
where
\begin{equation}
z=\frac{x_B+\sqrt{x_B^2-\rho_B^2}}{x_a+\sqrt{x_a^2-\rho_a^2}},
\label{eq:xa}
\end{equation}
and it is understood that $m_g=0$.
Using again the above bounds on $E_i$ ($i=b,g,B$), but imposing $z\le1$ instead
of $x_B\le x_a$, we now have
\begin{equation}
x_a^\text{min}=\frac{1}{2}\left(x_B+\sqrt{x_B^2-\rho_B^2}
+\frac{\rho_a^2}{x_B+\sqrt{x_B^2-\rho_B^2}}\right),
\end{equation}
while $x_b^\text{max}$ and $x_g^\text{max}$ go unchanged.
The kinematically allowed $x_B$ ranges now are
$\rho_B\le x_B\le E_B^\text{max}/E_b^\text{max}$ for $a=b$ and
$\rho_B\le x_B\le\tilde{E}_B^\text{max}/E_b^\text{max}$ for $a=g$.

Clearly, both Eqs.~(\ref{eq:master}) and (\ref{eq:lightcone}) may also be used
for $m_b=0$, to improve the ZM-VFN scheme by accommodating hadron-mass
corrections.
If also $m_B=0$ is put, then Eqs.~(\ref{eq:master}) and (\ref{eq:lightcone})
coincide reproducing the familiar factorization formula of the massless parton
model.

Taking a closer look at Eq.~(\ref{eq:master}), we observe that the GM-VFN
prediction implemented with the energy scaling variable $z=E_B/E_a$ is not
affected by finite-$m_B$ corrections inside the $x_B$ region accessible for
$m_B>m_b$.
This is quite different for the implementation with the light-cone-momentum
scaling variable $z=p_B^+/p_a^+$ via Eq.~(\ref{eq:lightcone}).
Of course, these observations carry over to the ZM-VFN case of $m_b=0$.

\boldmath
\section{Analytic results for $d\Gamma(t\to B+X)/d x_B$}
\label{sec:two}
\unboldmath

We now discuss the evaluation of the quantities
$d\hat{\Gamma}_a(\mu_R,\mu_F)/dx_a$ at NLO in the GM-VFN scheme.
Their counterparts in the FFN scheme contain the full $m_b$ dependence.
However, in the limit $m_b/m_t\to0$, they develop large logarithmic would-be
collinear singularities of the type $(\alpha_s/\pi)\ln(m_t^2/m_b^2)$, which
spoil the convergence of the perturbative expansion.
There is no conceptual necessity for FFs in this scheme.
They may still be introduced by hand, but there is no factorization theorem to
guarantee their universality. 

In the ZM-VFN scheme, where $m_b=0$ is put right from the start, such collinear
singularities are regularized by dimensional regularization in $D=4-2\epsilon$
space-time dimensions to become single poles in $\epsilon$, which are
subtracted at factorization scale $\mu_F$ and absorbed into the bare FFs
according to the modified minimal-subtraction ($\overline{\text{MS}}$) scheme.
This renormalizes the FFs, endowing them with $\mu_F$ dependence, and generates
in $d\hat{\Gamma}_a/dx_a$ finite terms of the form
$(\alpha_s/\pi)\ln(m_t^2/\mu_F^2)$, which are rendered perturbatively small
by choosing $\mu_F={\cal O}(m_t)$.
In this scheme, $m_b$ only sets the initial scale
$\mu_F^\text{ini}={\cal O}(m_b)$ of the DGLAP evolution, where ansaetze for the
$z$ dependences of the FFs $D_a(z,\mu_F^\text{ini})$ are injected.
The DGLAP evolution from $\mu_F^\text{ini}$ to $\mu_F$ then effectively resums
the problematic logarithms $(\alpha_s/\pi)\ln(m_t^2/m_b^2)$ of the FFN scheme.
However, all information on the $m_b$ dependence of $d\hat{\Gamma}_a/dx_a$ is
wasted.

The GM-VFN scheme is devised to resum the large logarithms in $m_b$ and to
retain the entire nonlogarithmic $m_b$ dependence at the same time.
This is achieved by introducing appropriate subtraction terms in the
NLO FFN expressions for $d\hat{\Gamma}_a/dx_a$, so that the NLO ZM-VFN results
are exactly recovered in the limit $m_b/m_t\to0$.
These subtraction terms are universal and so are the FFs in this scheme, as is
guaranteed by Collin's factorization theorem \cite{Collins:1998rz}.
If the same experimental data are fitted in the ZM-VFN and GM-VFN schemes, the
resulting FFs will be somewhat different.

For the sake of a comparative analysis of the GM-VFN and ZM-VFN schemes, we
need to know $d\hat{\Gamma}_a/dx_a$ at NLO in both schemes.
The NLO expressions for $d\hat{\Gamma}_b/dx_b$ in the ZM-VFN and FFN schemes
may be found in Eqs.~(5) and (6) and Eq.~(A.10) of Ref.~\cite{Corcella:2001hz},
respectively.
We verified them by an independent calculation and also derived those for
$d\hat{\Gamma}_g/dx_g$.
For the reader's convenience, we list here all our results.
In the ZM-VFN scheme, we have
\begin{eqnarray}
\frac{1}{\Gamma_0}\,\frac{d\hat{\Gamma}_b}{dx_b}&=&
\delta(1-x_b)+\frac{\alpha_s(\mu_R)}{2\pi}C_F
\left\{\left(\frac{1+x_b^2}{1-x_b}\right)_+
\left[\ln\frac{x_b^2(1-w)^2m_t^2}{\mu_F^2}+1\right]
+\frac{4x_b}{(1-x_b)_+}
\right.\nonumber\\
&&{}\times\left\{{{w(1-w)(1-x_b)^2}\over{(1+2w)[1-x_b(1-w)]}}-1\right\}
+2(1+x_b^2)\left(\frac{\ln(1-x_b)}{1-x_b}\right)_+
-\delta(1-x_b)
\nonumber\\
&&{}\times\left.\left[4\Li_2(w)+2\ln w\ln(1-w)+\frac{2w}{1-w}\ln w
+\frac{5+4w}{1+2w}\ln(1-w)
+\frac{15}{2}\right]\right\},
\nonumber\\
\frac{1}{\Gamma_0}\,\frac{d\hat{\Gamma}_g}{dx_g}&=&
\frac{\alpha_s(\mu_R)}{2\pi}C_F\left\{
\frac{1+(1-x_g)^2}{x_g}
\ln\frac{x_g^2(1-x_g)^2(1-w)^2m_t^2}{[1-x_g(1-w)]\mu_F^2}
\right.\nonumber\\
&&{}+\frac{1}{2(1+2w)[1-x_g(1-w)]^2}
\left[-(1-6w)(1-w)^2x_g^3+2(1-w)(1-2w)
\vphantom{\frac{1}{x_g}}
\right.\nonumber\\
&&{}\times\left.\left.(3+w)x_g^2
-(13-25w^2+6w^3)x_g+4(1+2w)(3-2w)-4\frac{1+2w}{x_g}
\right]\right\},
\label{eq:zmvfn}
\end{eqnarray}
where
\begin{equation}
\Gamma_0=\frac{G_Fm_t^3|V_{tb}|^2}{8\pi\sqrt2}(1-w)^2(1+2w)
\end{equation}
is the total decay width at LO, $G_F$ is Fermi's constant, $w=m_W^2/m_t^2$,
$C_F=(N_c^2-1)/(2N_c)=4/3$ for $N_c=3$ quark colors, and
$\Li_2(x)=-\int_0^x(dt/t)\ln(1-t)$ is the Spence function.
Integrating $d\hat{\Gamma}_b/dx_b$ of Eq.~(\ref{eq:zmvfn}) over $x_b$, we
recover the familiar result given by Eq.~(2.8) in connection with Eqs.~(3.1)
and (3.2) of Ref.~\cite{Jezabek:1988iv}.

In the FFN scheme, we have
\begin{eqnarray} 
{1\over{\Gamma_0}}\,{{d\hat{\Gamma}_b}\over{dx_b}}&=&\delta(1-x_b)+ 
{{C_F\alpha_s(\mu_R)}\over{\pi Q}}\left\{\left\{2s\left[
\Li_2\left( {{2Q}\over{1-s+Q}}\right)-
\Li_2\left( {{2Q}\over{s-b+Q}}\right)\right.\right.\right.
\nonumber\\
&&{}-\left.\ln(s+Q)\left(\ln{{1-s+Q}\over {\sqrt{w}}}+
\ln{{s-b+Q}\over{2s(1-\beta)}}\right)
+{1\over 2}\ln b\ln{{s-b+Q}\over{2s(1-\beta)}} \right]
\nonumber\\
&&{}+\left( 3{{Q^2}\over{G_0}}+s-b\right)\ln{{s+Q}\over \sqrt{b}}
+(1-b)\ln{{1-s+Q}\over {\sqrt{w}}}
\nonumber\\
&&{}+\left.Q\left[\left(6{{(w-b)(s-b)}\over{wG_0}}-1\right){{\ln b}\over 4}
-2\ln{{2s(1-\beta)}\over{\sqrt{w}}}-2
\right]\right\}\delta(1-x_b)
\nonumber\\
&&{}-2{\Phi (x_b)}\left[ {1\over {(1-x_b)_+}}+
{s\over{G_0}}\left(1+{{1+b}\over{2w}}\right)(1-x_b) - 1\right]
\nonumber\\
&&{}+\left. 2s\sqrt{x_b^2-\beta^2}\left[ 
2{s^2\over G_0}\, {{1-x_b}\over{1-2sx_b+b}} + 
{s\over{G_0}}\left(1+{{1+b}
\over{2w}}\right)(1-x_b) - 1\right]\right\}.
\nonumber\\
\frac{1}{\Gamma_0}\,\frac{d\hat{\Gamma}_g}{dx_g}&=&
\frac{C_F\alpha_s(\mu_R)s^2}{\pi wQG_0}\left\{
\left[(2s+3w)\frac{1+(1-x_g)^2}{x_g}-4\frac{b}{s}\,\frac{1-x_g}{x_g}\right]
\arcosh\frac{1-x_g}{\beta\sqrt{1-2sx_g}}
\right.\nonumber\\
&&{}+\frac{1-x_g}{(1-2sx_g)^2}\sqrt{1-\frac{\beta^2(1-2sx_g)}{(1-x_g)^2}}
\left[s^2(6s+w)x_g^2-2sx_g-wG_0
\right.\nonumber\\
&&{}\times\left.\left.\left(7sx_g-8+\frac{2}{sx_g}\right)\right]\right\},
\label{eq:ffn}
\end{eqnarray}
where
\begin{equation}
\Gamma_0=\frac{G_Fm_t^3|V_{tb}|^2}{2\pi\sqrt2}wQG_0
\end{equation}
is the total decay width at LO and, in the notation of
Ref.~\cite{Corcella:2001hz},
\begin{eqnarray}
b&=&\frac{m_b^2}{m_t^2},\qquad
s=\frac{1}{2}(1+b-w),\qquad
\beta=\frac{\sqrt b}{s},\qquad
Q=s\sqrt{1-\beta^2},
\nonumber\\
G_0&=&\frac{1}{2}\left[1+b-2w+\frac{(1-b)^2}{w}\right],\qquad
\Phi(x_b)=s\left(\sqrt{x_b^2-\beta^2}
-\artanh\frac{\sqrt{x_b^2-\beta^2}}{x_b}\right).\quad
\end{eqnarray}
Integrating $d\hat{\Gamma}_b/dx_b$ of Eq.~(\ref{eq:ffn}) over $x_b$, we recover
the familiar result given by Eq.~(2.8) in connection with Eqs.~(2.2)--(2.4) and
(2.6) of Ref.~\cite{Jezabek:1988iv}.

As explained above, the GM-VFN results are obtained by matching the FFN ones to
the ZM-VFN ones by subtraction, as
\begin{equation}
\left(\frac{1}{\Gamma_0}\,\frac{d\hat{\Gamma}_a}{dx_a}\right)_\text{GM-VFN}
=\left(\frac{1}{\Gamma_0}\,\frac{d\hat{\Gamma}_a}{dx_a}\right)_\text{FFN}
-\left(\frac{1}{\Gamma_0}\,\frac{d\hat{\Gamma}_a}{dx_a}\right)_\text{sub},
\end{equation}
where the subtraction terms are constructed as
\begin{equation}
\left(\frac{1}{\Gamma_0}\,\frac{d\hat{\Gamma}_a}{dx_a}\right)_\text{sub}
=\lim_{m_b\to0}
\left(\frac{1}{\Gamma_0}\,\frac{d\hat{\Gamma}_a}{dx_a}\right)_\text{FFN}
-\left(\frac{1}{\Gamma_0}\,\frac{d\hat{\Gamma}_a}{dx_a}\right)_\text{ZM-VFN}.
\end{equation}
Taking the limit $m_b\to0$ in Eq.~(\ref{eq:ffn}), we recover
Eq.~(\ref{eq:zmvfn}) up to the terms
\begin{eqnarray}
\left(\frac{1}{\Gamma_0}\,\frac{d\hat{\Gamma}_b}{dx_b}\right)_\text{sub}
&=&\frac{\alpha_s(\mu_R)}{2\pi}C_F
\left\{\frac{1+x_b^2}{1-x_b}\left[\ln\frac{\mu_F^2}{m_b^2}-2\ln(1-x_b)-1\right]
\right\}_+,
\label{eq:pff}\\
\left(\frac{1}{\Gamma_0}\,\frac{d\hat{\Gamma}_g}{dx_g}\right)_\text{sub}
&=&\frac{\alpha_s(\mu_R)}{2\pi}C_F
\frac{1+(1-x_g)^2}{x_g}\left(\ln\frac{\mu_F^2}{m_b^2}-2\ln x_g-1\right).
\end{eqnarray}
As already observed in Ref.~\cite{Corcella:2001hz}, Eq.~(\ref{eq:pff})
coincides with the perturbative FF of the transition $b\to b$
\cite{Mele:1990cw}.

\boldmath
\section{Analytic results for
$d\Gamma(t\to B\ell^+\nu_\ell+X)$\break
$/(dx_B\,d\cos\theta)$}
\label{sec:three}
\unboldmath

\begin{figure}
\begin{center}
\includegraphics[width=0.5\linewidth]{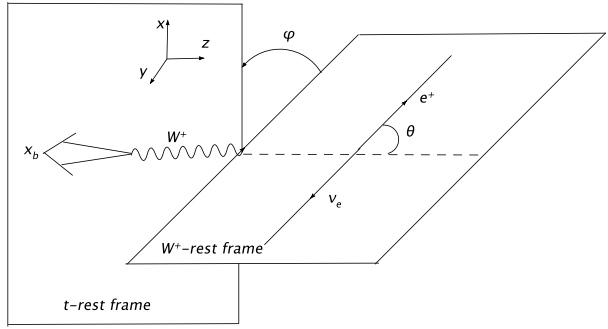}
\caption{\label{fig:angle}Definition of the polar angle $\theta$ in the
$W$-boson rest frame.}
\end{center}
\end{figure}

We now allow for the $W$ boson in process~(\ref{eq:proc}) to decay.
For definiteness, we consider its leptonic decay, which is cleaner than the
hadronic one, so that we are dealing with
the process
\begin{equation}
t(p_t)\to b(p_b)+W^+(p_W)(+g(p_g))\to B(p_B)+\ell^++\nu_\ell+X,
\label{eq:cascade}
\end{equation}
for which we wish to calculate the doubly differential partial decay width,
$d\Gamma/(dx_B\,d\cos\theta)$, at NLO.
As mentioned above, $\theta$ is the decay angle of the charged lepton $\ell^+$
in the $W$-boson rest frame (see Fig.~\ref{fig:angle}).

As will be demonstrated in Sec.~\ref{sec:four}, the finite-$m_b$ corrections
are rather small in the case of process~(\ref{eq:proc}), much smaller than the
contribution from gluon fragmentation.
In the following, we, therefore, set $m_b=0$, {\it i.e.}\ we  work in the
ZM-VFN scheme, but we retain $m_B\ne0$ and gluon fragmentation.
Furthermore, we put $p_W^2=m_W^2$, {\it i.e.}\ we work in the narrow-width
approximation.
This allows us to employ the helicity density matrix formalism, so that the
squared decay amplitude of process~(\ref{eq:cascade}) factorizes into the
squared amplitude of process~(\ref{eq:proc}), with definite $W$-boson
polarization, and the squared amplitude of the subsequent decay
$W^+\to\ell^++\nu_\ell$ of the $W$ boson just so polarized.
The three degrees of massive-vector-boson polarization, $\lambda=0,\pm1$, are
conveniently implemented by applying the covariant projection operators
\begin{eqnarray} 
\varepsilon^{\mu}(0)\varepsilon^{\nu*}(0)&=&\frac{w}{\vec{p}_W^{\;2}}
\left(p_t^\mu-\frac{p_t\cdot p_W}{m_W^2}p_W^\mu\right)
\left(p_t^\nu-\frac{p_t\cdot p_W}{m_W^2}p_W^\nu\right),
\nonumber\\
\varepsilon^{\mu}(\pm)\varepsilon^{\nu*}(\pm)&=&
\frac{1}{2}\left[-g^{\mu\nu}+\frac{p_W^{\mu}p_W^{\nu}}{m_W^2}
-\frac{w}{\vec{p}_W^{\;2}}
\left(p_t^\mu-\frac{p_t\cdot p_W}{m_W^2}p_W^\mu\right)
\left(p_t^\nu-\frac{p_t\cdot p_W}{m_W^2}p_W^\nu\right)
\right.\nonumber\\
&&{}\mp\left.\frac{i\epsilon^{\mu\nu\rho\sigma}}{m_t|\vec{p}_W|}
(p_t)_\rho(p_W)_\sigma\right],
\label{eq:pol}
\end{eqnarray}
where $\epsilon^{0123}=1$ and
$|\vec{p}_W|=\sqrt{(p_t\cdot p_W/m_t)^2-m_W^2}=m_t\sqrt{[1-s(x_b+x_g)]^2-w}$
is the modulus of the $W$-boson three-momentum in the top-quark rest frame.
Performing the polarization sum, we recover the familiar completeness relation
\begin{equation}
\sum_{\lambda=-1}^{+1}\varepsilon^{\mu}(\lambda)\varepsilon^{\nu*}(\lambda)
=-g^{\mu\nu}+\frac{p_W^{\mu}p_W^{\nu}}{m_W^2},
\label{eq:unpol}
\end{equation}
used in Sec.~\ref{sec:two}.

Repeating the calculation of Sec.~\ref{sec:two} at NLO in the ZM-VFN scheme
with Eq.~(\ref{eq:pol}) instead of Eq.~(\ref{eq:unpol}), we obtain
\begin{equation}
\frac{d^2\hat{\Gamma}_a}{dx_a\,d\cos\theta}
=\frac{3}{8}(1+\cos\theta)^2\frac{d\hat{\Gamma}_a^+}{dx_a}
+\frac{3}{8}(1-\cos\theta)^2\frac{d\hat{\Gamma}_a^-}{dx_a}
+\frac{3}{4}\sin^2\theta\frac{d\hat{\Gamma}_a^0}{dx_a},
\label{eq:angular}
\end{equation}
where $d\hat{\Gamma}_a^\lambda/dx_a$ are $\theta$-independent coefficient
functions listed in Eq.~(\ref{eq:app}) of the Appendix. 
As in Sec.~\ref{sec:two}, the top quark is assumed to be unpolarized.
At LO, $\lambda=0$ refers to the case when the top-quark spin is passed on to
the bottom quark as is, while it is flipped for $\lambda=-1$;
$\lambda=+1$ is prohibited by angular-momentum conservation, so that
$d\hat{\Gamma}_a^+/dx_a$ vanishes at LO.
At NLO, all values of $\lambda$ are allowed because of the presence of the
additional spin-one boson $g$.

There are two powerful checks for the correctness of Eq.~(\ref{eq:app}).
Firstly, integrating Eq.~(\ref{eq:angular}) over $\cos\theta$, we obtain
\begin{equation}
\frac{d\hat{\Gamma}_a}{dx_a}
=\sum_{\lambda=-1}^{+1}\frac{d\hat{\Gamma}_a^\lambda}{dx_a},
\end{equation}
which agrees with Eq.~(\ref{eq:zmvfn}) upon insertion of our expressions for
$d\hat{\Gamma}_a^\lambda/dx_a$.
Secondly, integrating $d\hat{\Gamma}_b^\lambda/dx_b$ over $x_b$, we recover the
results presented in Eqs.~(15)--(17) of Ref.~\cite{Fischer:2000kx}.

The structure of the angular dependence of Eq.~(\ref{eq:angular}) is preserved
by the convolution with the FFs according to Eq.~(\ref{eq:master}) [or
Eq.~(\ref{eq:lightcone})], and we may project out the hadronic counterparts of
$d\hat{\Gamma}_a^\lambda/dx_a$,
\begin{equation}
\frac{d\Gamma^\lambda}{dx_B}=\sum_{a=b,g}\int_{x_a^\text{min}}^{x_a^\text{max}}
\frac{dx_a}{x_a}\,\frac{d\hat{\Gamma}_a^\lambda}{dx_a}(\mu_R,\mu_F)
D_a\left(\frac{x_B}{x_a},\mu_F\right),
\end{equation}
from the measured $\theta$ dependence of $d^2\Gamma/(dx_B\,d\cos\theta)$ as
\begin{equation}
\frac{d\Gamma^\lambda}{dx_B}=\int_{-1}^1d\cos\theta\,f^\lambda(\cos\theta)
\frac{d^2\Gamma}{dx_B\,d\cos\theta},
\end{equation}
where
\begin{equation}
f^{\pm1}(x)=-\frac{1}{2}\pm x+\frac{5}{2}x^2,\qquad
f^0(x)=2-5x^2.
\end{equation}
In this way, we obtain three independent $x_B$ distributions, which we can use
to constrain the $B$-hadron FFs.

\section{Numerical analysis}
\label{sec:four}

We are now in a position to explore the phenomenological consequences of our
results by performing a numerical analysis.
We adopt from Ref.~\cite{Nakamura:2010zzi} the input parameter values
$G_F = 1.16637\times10^{-5}$~GeV$^{-2}$,
$m_W = 80.399$~GeV,
$m_t = 172.9$~GeV,
$m_b = 4.78$~GeV,
$m_B = 5.279$~GeV, and
$|V_{tb}|=0.999152$.
We evaluate $\alpha_s^{(n_f)}(\mu_R)$ at NLO in the $\overline{\text{MS}}$
scheme using Eq.~(9.5) of Ref.~\cite{Nakamura:2010zzi}, retaining only the
first two terms within the parentheses, with $n_f=5$ active quark flavors and
asymptotic scale parameter $\Lambda_{\overline{\text{MS}}}^{(5)}=231$~MeV
adjusted such that $\alpha_s^{(5)}(m_Z) = 0.1184$ for $m_Z = 91.1876$~GeV
\cite{Nakamura:2010zzi}.
We employ the nonperturbative $B$-hadron FFs that were determined at NLO in
the ZM-VFN scheme through a joint fit \cite{Kniehl:2008zza} to
$e^+e^-$-annihilation data taken by ALEPH \cite{Heister:2001jg} and OPAL
\cite{Abbiendi:2002vt} at CERN LEP1 and by SLD \cite{Abe:1999ki} at SLAC SLC.
Specifically, the power ansatz $D_b(z,\mu_F^\text{ini})=Nz^\alpha(1-z)^\beta$
was used as the initial condition for the $b\to B$ FF at
$\mu_F^\text{ini}=4.5$~GeV, while the gluon and light-quark FFs were generated
via the DGLAP evolution.
The fit yielded $N=4684.1$, $\alpha=16.87$, and $\beta=2.628$ with
$\chi_\text{d.o.f.}^2=1.495$.
We choose $\mu_R=\mu_F=m_t$.

\begin{figure}
\begin{center}
\includegraphics[width=0.8\linewidth,bb=37 193 550 628]{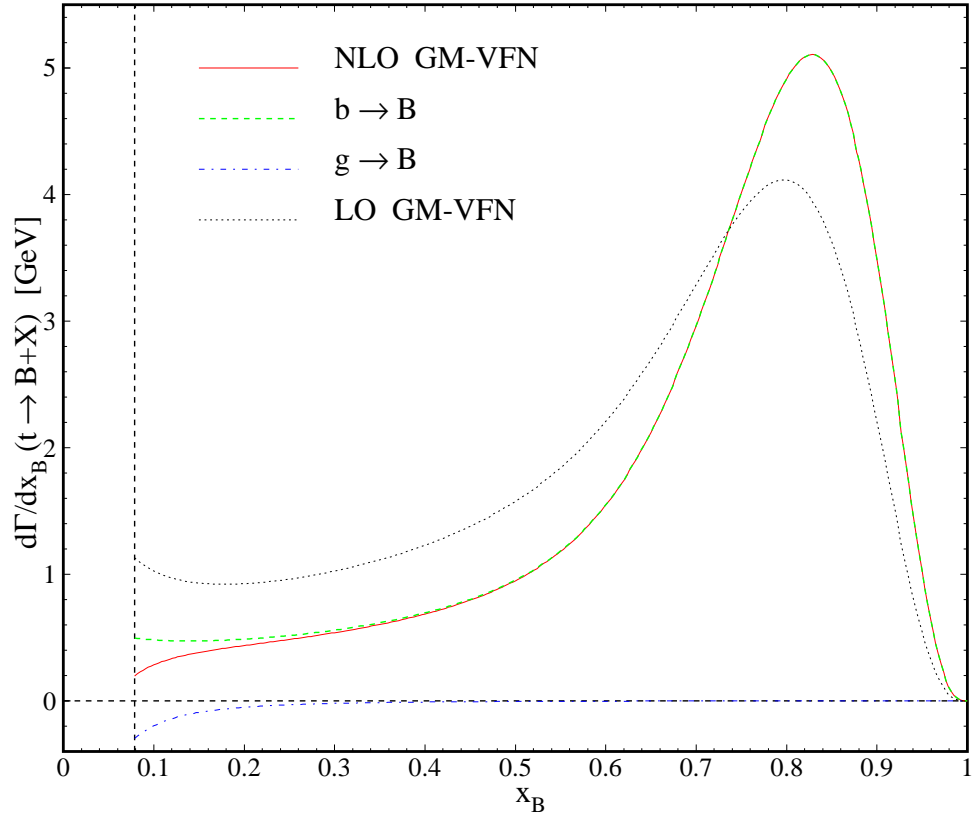}
\caption{\label{fig:two}%
$d\Gamma(t\to B+X)/dx_B$ as a function of $x_B$ in the GM-VFN ($m_b\ne0$)
scheme including finite-$m_B$ corrections implemented with the energy scaling
variable $z=E_B/E_a$ using Eq.~(\ref{eq:master}).
The NLO result (solid line) is compared to the LO one (dotted line) and broken
up into the contributions due to $b\to B$ (dashed line) and $g\to B$
(dot-dashed line) fragmentation.}
\end{center}
\end{figure}

We first consider the quantity $d\Gamma(t\to B+X)/dx_B$ taking the $W$ boson
to be stable.
Our most reliable prediction for it is made at NLO in the GM-VFN scheme and
includes finite-$m_B$ corrections.
For the time being, we implement the latter in terms of the energy scaling
variable $z=E_B/E_a$ using Eq.~(\ref{eq:master}).
In Fig.~\ref{fig:two}, we study for this prediction the size of the NLO
corrections, by comparing the LO (dotted line) and NLO (solid line) results,
and the relative importance of the $b\to B$ (dashed line) and $g\to B$
(dot-dashed line) fragmentation channels at NLO.
In order to expose the size of the NLO corrections at the parton level, we
evaluate the LO result using the same NLO FFs.
We observe from Fig.~\ref{fig:two} that the NLO corrections lead to a
significant enhancement of the partial decay width in the peak region and
above, by as much as 25\%, at the expense of a depletion in the lower-$x_B$
range.
Furthermore, the peak position is shifted towards higher values of $x_B$.
The $g\to B$ contribution is throughout negative and appreciable only in the
low-$x_B$ region, for $x_B\alt0.2$.
For higher values of $x_B$ the NLO result is practically exhausted by the
$b\to B$ contribution, as expected \cite{Corcella:2001hz}.
In the following, we only consider NLO results unless otherwise stated.

\begin{figure}
\begin{center}
\includegraphics[width=0.8\linewidth,bb=35 193 550 628]{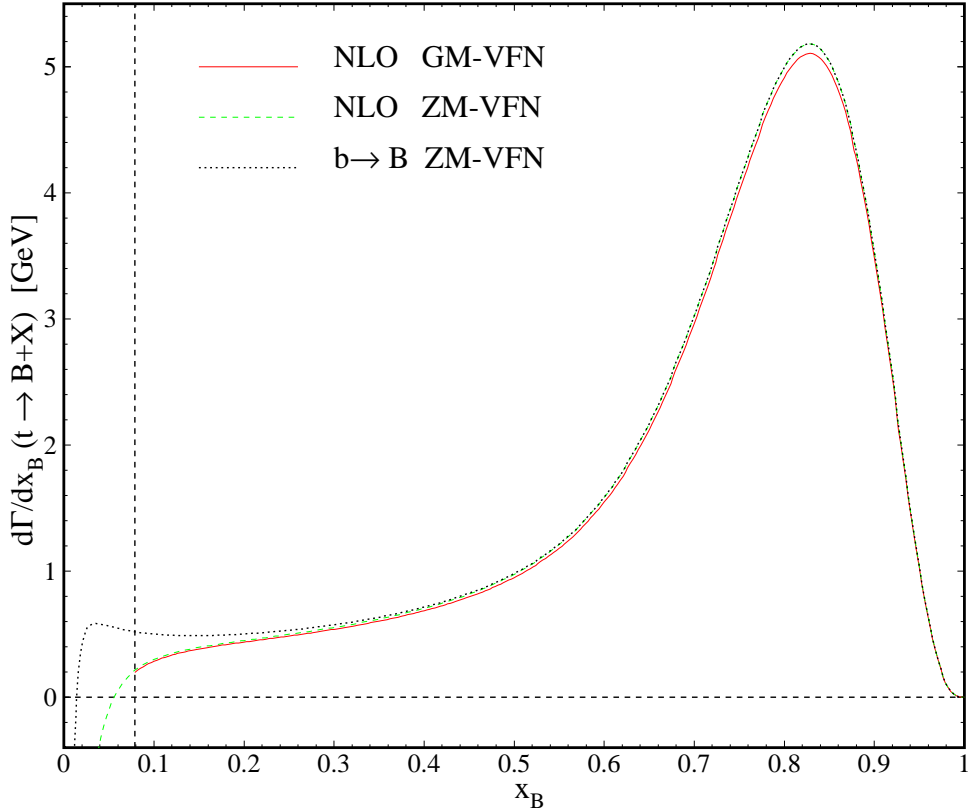}
\caption{\label{fig:three}%
$d\Gamma(t\to B+X)/dx_B$ as a function of $x_B$ at NLO.
The GM-VFN ($m_b\ne0$) result including finite-$m_B$ corrections implemented
with the energy scaling variable $z=E_B/E_a$ using Eq.~(\ref{eq:master})
(solid line) is compared to the ZM-VFN ($m_b=0$) results for $m_B=0$ excluding
(dotted line) and including (dashed line) $g\to B$ fragmentation.}
\end{center}
\end{figure}

In Fig.~\ref{fig:three}, we study the improvement over previous calculations,
at NLO in the ZM-VFN scheme neglecting $g\to B$ fragmentation and finite-$m_B$
corrections (dotted line), gained by switching to our preferred mode of
evaluation, at NLO in the GM-VFN scheme including $g\to B$ fragmentation and
finite-$m_B$ corrections (solid line).
For comparison, we also show the full ZM-VFN prediction, including $g\to B$
fragmentation, for $m_B=0$ (dashed line).
We observe from Fig.~\ref{fig:three} that the improvement is threefold.
First, the finite-$m_B$ corrections are responsible for the appearance of the
threshold at $x_B=\rho_B=0.07783$.
Second, $g\to B$ fragmentation leads to a significant reduction in size in the
threshold region, as is familiar from Fig.~\ref{fig:two}. 
Third, the finite-$m_b$ corrections lead to a moderate reduction in size
throughout the whole $x_B$ range allowed.
A similar observation was made for the $x_b$ distribution in the
perturbative-FF approach \cite{Corcella:2001hz}.

\begin{figure}
\begin{center}
\includegraphics[width=0.8\linewidth,bb=19 190 550 628]{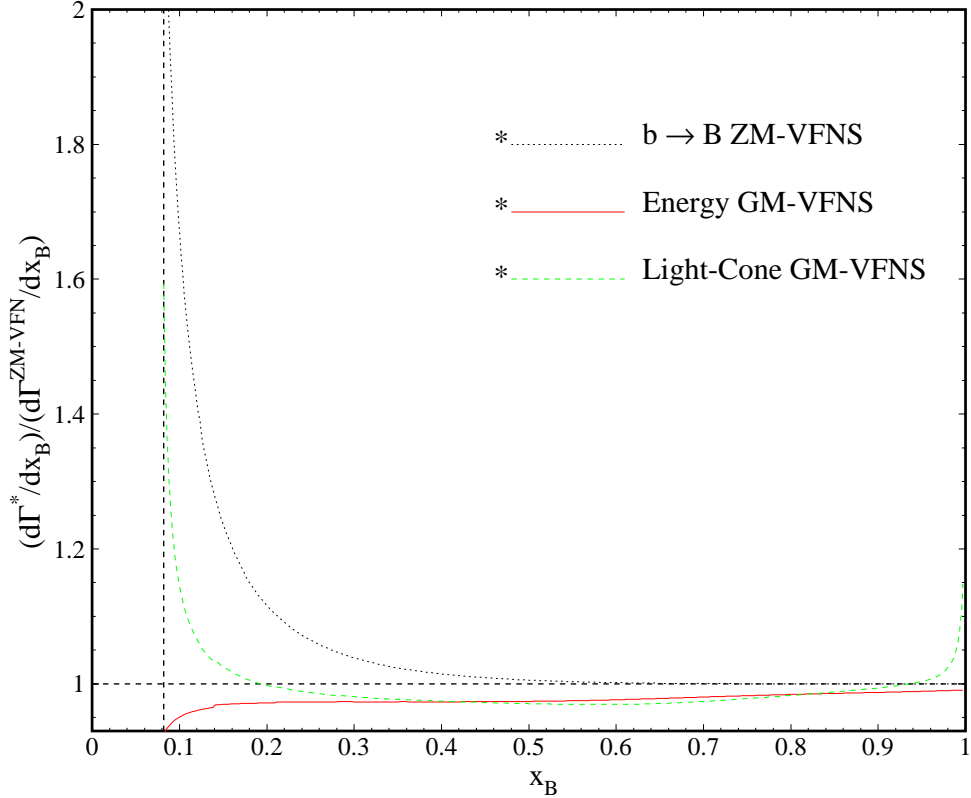}
\caption{\label{fig:four}%
$d\Gamma(t\to B+X)/dx_B$ as a function of $x_B$ at NLO in the GM-VFN
($m_b\ne0$) scheme including finite-$m_B$ corrections.
The implementation with the energy scaling variable $z=E_B/E_a$ via
Eq.~(\ref{eq:master}) (solid line) is compared to the one with the
light-cone-momentum scaling variable $z=p_B^+/p_a^+$ via
Eq.~(\ref{eq:lightcone}) (dashed line).
For comparison, also the ZM-VFN ($m_b=0$) result for $m_B=0$ excluding
$g\to B$ fragmentation is shown (dotted line).
All results are normalized to the ZM-VFN ($m_b=0$) one for $m_B=0$ including
$g\to B$ fragmentation.}
\end{center}
\end{figure}

For a more quantitative interpretation of Fig.~\ref{fig:three}, we consider in
Fig.~\ref{fig:four} the ZM-VFN result for $m_B=0$ without $g\to B$
fragmentation (dotted line) and the full GM-VFN result implemented with the
energy scaling variable (solid line), both normalized to the full ZM-VFN result
for $m_B=0$.
We observe that the omission of $g\to B$ fragmentation entails an excess by a
factor of up to 2 close to threshold, while the finite-$m_b$ corrections just
amount to a few percent.
This motivates us to adopt the ZM-VFN scheme including $g\to B$ fragmentation
in the treatment of
$d^2\Gamma(t\to B\ell^+\nu_\ell+X)/(dx_B\,d\cos\theta)$ below.

Figure~\ref{fig:four} also includes the result obtained in the GM-VFN scheme
implemented with the light-cone-momentum scaling variable $z=p_B^+/p_a^+$ via
Eq.~(\ref{eq:lightcone}) (dashed line).
Comparison with its counterpart for the energy scaling variable allows us to
assess the theoretical uncertainty in the implementation of finite-$m_B$
corrections in the GM-VFN scheme.
We observe from Fig.~\ref{fig:four} that the two implementations of
finite-$m_B$ corrections lead to very similar results over most of the $x_B$
range.
Appreciable differences only appear close to the kinematical bounds.
However, these should be taken with a grain of salt because the FFs used here
\cite{Kniehl:2008zza} refer to the ZM-VFN scheme and do not carry any specific
information on finite-$m_B$ effects and the scheme of their implementation.

\begin{figure}
\begin{center}
\includegraphics[width=0.8\linewidth,bb=18 190 550 628]{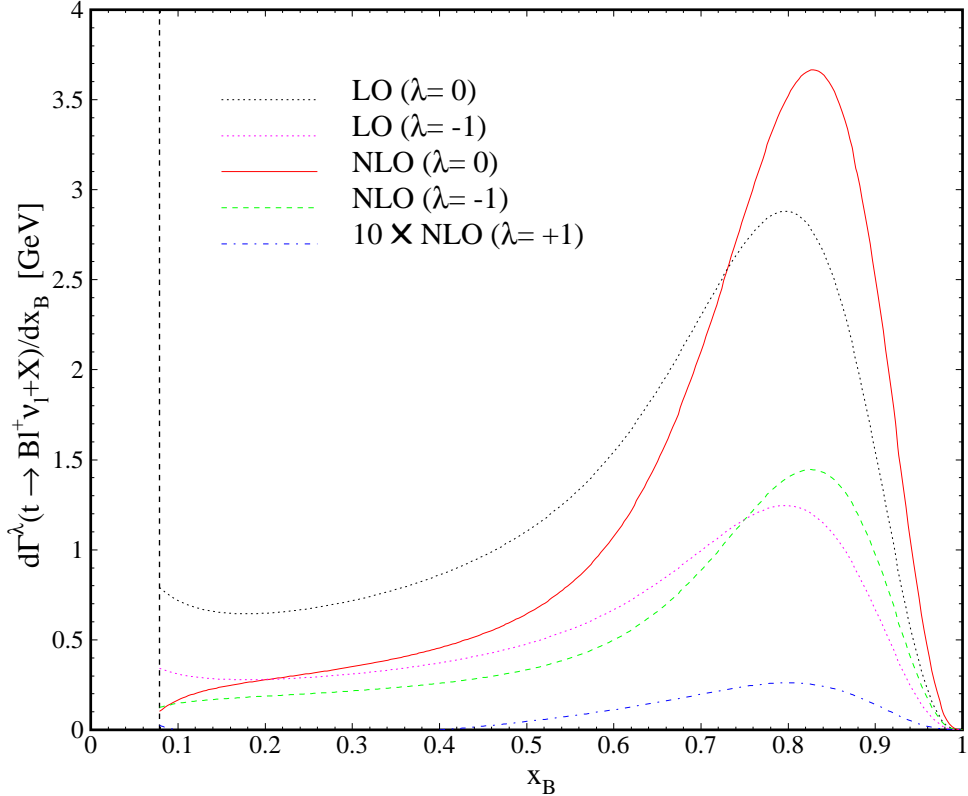}
\caption{\label{fig:five}%
$d\Gamma^\lambda(t\to B\ell^+\nu_\ell+X)/dx_B$ as functions of $x_B$ in the
ZM-VFN ($m_b=0$) scheme including finite-$m_B$ corrections implemented with the
energy scaling variable $z=E_B/E_a$ using Eq.~(\ref{eq:master}).
The NLO results for $\lambda=0$ (solid line), $\lambda=-1$ (dashed line),
and $\lambda=+1$ (dot-dashed line) and the LO results for $\lambda=0$ (upper
dotted line) and $\lambda=-1$ (lower dotted line) are compared to each other.}
\end{center}
\end{figure}

In the remainder of this section, we consider the quantity
$d^2\Gamma(t\to B\ell^+\nu_\ell+X)$
$/(dx_B\,d\cos\theta)$, which we study at
NLO in the ZM-VFN scheme including $g\to B$ fragmentation and finite-$m_B$
corrections evaluated using the energy scaling variable.
In Fig.~\ref{fig:five}, we present the $x_B$ distributions of
$d\Gamma^\lambda/dx_B$ for $\lambda=0$ (solid line), $\lambda=-1$ (dashed
line), and $\lambda=+1$ (dot-dashed line).
For comparison, also the LO results for $\lambda=0,-1$ (dotted lines) are
shown.
As may be gleaned from Eq.~(\ref{eq:app}), the latter amount to respectively
$1/(1+2w)=69.59\%$ and $2w/(1+2w)=30.41\%$ of the LO result.
In these two cases, the NLO corrections are similar in size and shape to the
unpolarized case (see Fig.~\ref{fig:two}).
As explained in Sec.~\ref{sec:three}, $\lambda=+1$ is prohibited at LO, which
explains the smallness of the corresponding result.

\section{Conclusions}
\label{sec:five}

Let alone its discovery potential with regard to the Higgs boson and new
physics beyond the standard model, the LHC is a formidable top factory, which,
among other things, will allow for the study of the dominant decay mode
$t\to BW^++X$ with unprecedented precision in the long run.
In particular, this will enable us to deepen our understanding of the
nonperturbative aspects of $B$-hadron formation by hadronization and to pin
down the $b\to B$ and $g\to B$ FFs.
The key observable for this purpose is the $x_B$ distribution $d\Gamma/dx_B$ of
$t\to B+X$.
By measuring the angular distribution of the $W$-boson decay products, also
the three components $d\Gamma^\lambda/dx_B$ of $d\Gamma/dx_B$ corresponding to
the polarization states $\lambda=0,\pm1$ of the $W$ boson may be determined,
which constrain these FFs even further.

We studied the quantity $d\Gamma/dx_B$ at NLO in the GM-VFN scheme
\cite{Kneesch:2007ey,Kramer:2001gd,Kramer:2003jw,Kniehl:2004fy,Kniehl:2008zza}
using nonperturbative $B$-hadron FFs determined by a global fit
\cite{Kniehl:2008zza} of experimental data from the $Z$ factories
\cite{Heister:2001jg,Abbiendi:2002vt,Abe:1999ki}, relying on their universality
and scaling violations \cite{Collins:1998rz}.
This allowed us to investigate for the first time finite-$m_b$ corrections and
the contribution from gluon fragmentation to $d\Gamma/dx_B$.
We also analyzed the size of finite-$m_B$ effects and their theoretical
uncertainty due to the freedom in the choice of the scaling variable.
Since the finite-$m_b$ effects turned out to so moderate, we neglected them,
for simplicity, in our study of $d\Gamma^\lambda/dx_B$, which we treated at NLO
in the ZM-VFN scheme taking into account finite-$m_B$ effects.

Comparing future measurements of $d\Gamma/dx_B$ and $d\Gamma^\lambda/dx_B$ at
the LHC with our NLO predictions, we will be able to test the universality and
scaling violations of the $B$-hadron FFs.
These measurements will ultimately be the primary source of information on the
$B$-hadron FFs.
The formalism elaborated here is also applicable to the production of hadron
species other than $B$ hadrons, {\it e.g.}\ pions, kaons, protons, $D$ mesons,
{\it etc.}, through top-quark decay.

\section*{Acknowledgment}

We thank Kirill Melnikov for a useful communication regarding
Ref.~\cite{Corcella:2009rs} and Elena Scherbakova for technical assistance.
This work was supported in part by the German Federal Ministry for Education
and Research BMBF through Grant No.\ 05~HT6GUA and by the Helmholtz
Association HGF through Grant No.\ Ha~101.
The work of S.M.M.N. was supported in part by the Ministry of Science,
Research, and Technology of Iran.

\section*{Appendix}

The coefficient functions $\hat{\Gamma}_a^\lambda$ in Eq.~(\ref{eq:angular})
exhibit the structure
\begin{eqnarray}
\frac{1}{\Gamma_0}\,\frac{d\hat{\Gamma}_a^0}{dx_a}
&=&\frac{1}{1+2w}\left[\delta_{ab}\delta(1-x_a)
+\frac{\alpha_s(\mu_R)}{2\pi}\left(P_{ab}(x_a)\ln\frac{m_t^2}{\mu_F^2}
+C_FC_a^0(x_a)\right)\right],
\nonumber\\
\frac{1}{\Gamma_0}\,\frac{d\hat{\Gamma}_a^-}{dx_a}
&=&\frac{2w}{1+2w}\left[\delta_{ab}\delta(1-x_a)
+\frac{\alpha_s(\mu_R)}{2\pi}\left(P_{ab}(x_a)\ln\frac{m_t^2}{\mu_F^2}
+C_FC_a^-(x_a)\right)\right],
\nonumber\\
\frac{1}{\Gamma_0}\,\frac{d\hat{\Gamma}_a^+}{dx_a}
&=&\frac{2w}{1+2w}\,\frac{\alpha_s(\mu_R)}{2\pi}C_FC_a^+(x_a),
\label{eq:app}
\end{eqnarray}
where
\begin{eqnarray}
P_{qq}(x)&=&C_F\left(\frac{1+x^2}{1-x}\right)_+,
\nonumber\\
P_{gq}(z)&=&C_F\frac{1+(1-x)^2}{x},
\end{eqnarray}
are the timelike $q\to q$ and $q\to g$ splitting functions at LO, respectively.
Introducing the short-hand notation
\begin{equation}
w_\pm=1\pm\sqrt{w},
\end{equation}
and defining
\begin{eqnarray}
L_1&=&\ln(1-x)\theta\left(\frac{1}{w_+}-x\right)
+\ln\left(\sqrt{w}x\right)\theta\left(x-\frac{1}{w_+}\right),
\nonumber\\
L_2&=&2\ln\frac{1-x(1-w)}{\sqrt{w}}
\theta\left(x-\frac{1}{w_+}\right),
\end{eqnarray}
where $\theta(x)=\int_{-\infty}^xdt\,\delta(t)$ is the Heaviside step function,
we have
\begin{eqnarray}
C_b^0(x)&=&-\delta(1-x)
\left[4\Li_2(w)+2\ln w\ln(1-w)+\frac{2w}{1-w}\ln w
+3\frac{2+5w}{1+2w}\right]
+\frac{1}{(1-x)_+}
\nonumber\\
&&{}\times\left\{2(1+x^2)\ln[x(1-w)]
-2\frac{1+4w}{1+2w}\right\}
+2(1+x^2)\left[\frac{\ln(1-x)}{1-x}\right]_+
-4w B_1
\nonumber\\
&&{}-\frac{8w[4-x(3+w)]}{(1-w)(2-xw_-)(2-xw_+)}
+\frac{1}{1+2w}\left(\frac{3+13w+8w^2}{1-w}-x\right),
\nonumber\\
C_b^-(x)&=&-\delta(1-x)\left[4\Li_2(w)+2\ln w\ln(1-w)
+\frac{2w}{1-w}\ln w
+\frac{1-w}{w}\ln(1-w)
\right.
\nonumber\\
&&{}+\left.\frac{3(7+18w)}{4(1+2w)}\right]
+\frac{1}{(1-x)_+}\left[2(1+x^2)\ln[x(1-w)]
-x^3\sqrt{\frac{1-w}{(2-xw_-)(2-xw_+)}}
\right.
\nonumber\\
&&{}-\left.
\vphantom{\sqrt{\frac{1-w}{(2-xw_-)(2-xw_+)}}}
\frac{4w}{1+2w}\right]
+2(1+x^2)\left[\frac{\ln(1-x)}{1-x}\right]_+
-\left(\frac{1+w}{1-w}-x\right)
\{L_2-\ln[1-x(1-w)]\}
\nonumber\\
&&{}+B_1
-\frac{1}{1-x}\left[B_2
-\frac{x(2-7x+4x^2)}{2(2-xw_-)(2-xw_+)}\right]
+\frac{1-x}{1-x(1-w)}
+\frac{3+x}{2}
-\frac{1+x}{1+2w},
\nonumber\\
C_b^+(x)&=&\frac{3}{4}\delta(1-x)
+\frac{x^3}{(1-x)_+}\sqrt{\frac{1-w}{(2-xw_-)(2-xw_+)}}
+\left(\frac{1+w}{1-w}-x\right)
\nonumber\\
&&{}\times\{L_2-\ln[1-x(1-w)]\}
+B_1+\frac{1}{1-x}\left[B_2
+\frac{x(6-9x+4x^2)}{2(2-xw_-)(2-xw_+)}\right]
\nonumber\\
&&{}+\frac{(1-x)[1+x(1-w)]}{2[1-x(1-w)]},
\end{eqnarray}
with
\begin{eqnarray}
B_1&=&\frac{1}{2\sqrt{w}}\left[
\frac{w_+(1-xw_-)^2\left(2\sqrt{w}+xw_-^2\right)}
{w_-(2-xw_-)^2}\ln(1-xw_-)
+\frac{w_-(1-xw_+)^2\left(2\sqrt{w}-xw_+^2\right)}
{w_+(2-xw_+)^2}
\right.\nonumber\\
&&{}\times\left.
\vphantom{\frac{w_-(1-xw_+)^2\left(2\sqrt{w}-xw_+^2\right)}
{w_+(2-xw_+)^2}}
\ln|1-xw_+|\right],
\nonumber\\
B_2&=&(1+x^2)L_1
+\sqrt{\frac{1-w}{(2-xw_-)(2-xw_+)}}
\nonumber\\
&&{}\times\left\{
\left[\frac{x^2(2-x)^2}{4(2-xw_-)(2-xw_+)}
-\frac{x(1+x^2)}{2}
-\frac{x(12-11x)}{4(1-w)}
+\frac{1-x}{(1-w)^2}
\right]\right.
\nonumber\\
&&\times\left.
\ln{\textstyle\frac{2-x\{1-3w+[2-x(1-w)]^2\}
+(1-xw_-)|1-xw_+|\sqrt{(1-w)(2-xw_-)(2-xw_+)}}
{2-x(1+w)+\sqrt{(1-w)(2-xw_-)(2-xw_+)}}}-x^3\right\}
\nonumber\\
&&-\frac{(1-xw_-)|1-xw_+|[2-x(1+w)]}
{(1-w)(2-xw_-)(2-xw_+)}
+\frac{1}{2(1-w)},
\end{eqnarray}
and
\begin{eqnarray}
C_g^0(x)&=&\frac{1+(1-x)^2}{x}\left\{2\ln[(1-w)x(1-x)]
+\frac{1}{1+2w}\right\}
-4w G_1
\nonumber\\
&&{}+\frac{1-x}{2x}\left\{
\frac{1-x}{[1-x(1-w)]^2}
-\frac{2(1+x)}{1-x(1-w)}
+3(3+x)
-\frac{16}{1-w}
\right\}
\nonumber\\
&&{}-\left[x
-\frac{2(1+5w-2w^2)}{1-w}
+\frac{2(1+12w+3w^2)}{x(1-w)^2}
-\frac{16w(1+w)}{x^2(1-w)^3}
\right]\ln[1-x(1-w)],
\nonumber\\
C_g^-(x)&=&2\frac{1+(1-x)^2}{x}
\left\{\ln[(1-w) x(1-x)]
+\frac{1+w}{1+2w}\right\}
+G_1-G_2
\nonumber\\
&&{}-\frac{1}{1-w}\left[
3
-\frac{7+6w-2w^2}{x(1-w)}
+\frac{4(1+2w)}{x^2(1-w)^2}\right]
\ln[1-(1-w)x]
\nonumber\\
&&{}
-\frac{w[2-x(1-w)(2+w)]}{4(1-w)[1-x(1-w)]^2}
-\frac{x}{4}
+\frac{4-7w}{2(1-w)}
-\frac{5-19w}{4x(1-w)},
\nonumber\\
C_g^+(x)&=&G_1+G_2
-\left[x
+\frac{1}{1-w}
-\frac{5}{x(1-w)^2}
+\frac{4}{x^2(1-w)^3}\right]\ln[1-(1-w)x]
\nonumber\\
&&{}-\frac{w[2-x(1-w)(2+w)]}{4(1-w)[1-x(1-w)]^2}
+\frac{3}{4}x
-\frac{4-w}{2(1-w)}
+\frac{5-3w}{4x(1-w)},
\end{eqnarray}
with
\begin{eqnarray}
G_1&=&\frac{1}{2x^2\sqrt{w}}\left[
\frac{w_+(1-xw_-)^2(2-xw_-^2)}{w_-^3}\ln(1-xw_-)
-\frac{w_-(1-xw_+)^2(2-xw_+^2)}{w_+^3}
\right.
\nonumber\\
&&{}\times\left.
\vphantom{\frac{w_+(1-xw_-)^2(2-xw_-^2)}{w_-^3}}
\ln|1-xw_+|
\right],
\nonumber\\
G_2&=&\frac{1+(1-x)^2}{x}L_1
+\left[\frac{x}{2}
-\frac{1}{1-w}
+\frac{1+3w-w^2}{x(1-w)^2}
-\frac{4w}{x^2(1-w)^3}\right]L_2
\nonumber\\
&&{}+\frac{(1-xw_-)|1-xw_+|}{4[1-x(1-w)]^2}
\left[x(1-w)
-2(2+w)
+\frac{7(1+w)}{x(1-w)}
-\frac{4(1+w)}{x^2(1-w)^2}\right]
\nonumber\\
&&{}+\frac{1+w}{x^2(1-w)^2}.
\end{eqnarray}

\end{document}